# A Computational Optimisation Study of Hip Implant Using Density Mapping Functionally Graded Biomimetic TPMS-based Lattice Structures

Mahtab Vafaeefar[1], Conall Quinn[1], Kevin M. Moerman[2,3], Ted J. Vaughan[1]

[1] Biomechanics Research Centre (BMEC), School of Engineering, Institute for Health Discovery and Innovation, College of Science and Engineering, University of Galway, Ireland.

[2] Mechanical Engineering, School of Engineering and Informatics, College of Science and Engineering, University of Galway, Ireland.

[3] Griffith Centre of Biomedical and Rehabilitation Engineering (GCORE), Griffith University, Gold Coast, Australia.

*Address for correspondence:*
Prof Ted J Vaughan
Professor in Biomedical Engineering,
Biomechanics Research Centre (BMEC)
Institute for Health Discovery and Innovation,
College of Science and Engineering,
University of Galway
Galway
Ireland
Phone: (+353) 91-493084
Email: ted.vaughan@universityofgalway.ie

**Keywords**
Hip Implant, Optimisation, Functionally Graded Lattices, Finite Element Modelling, Bone Remodelling, Density Mapping Lattice

# Abstract

This study presents a computational optimisation framework of a hip implant through the development of a functionally graded biomimetic lattice structure, whose design was structurally optimised to limit stress shielding. The optimisation technique was inspired by the inverse of a bone remodelling algorithm, promoting an even stress distribution throughout the design region, by reducing the density and consequently the stiffness, in regions where strain energy was higher than the reference level. The result of the optimisation technique provided a non-uniform graded density distribution field that showed lower density level on the sides of the implant stem, and higher material density around the medial axis of the stem. The optimised material distribution was captured using mapping of a triply periodic minimal surface lattice structure on the implant, which resulted in porous lattice surfaces inside the solid implant. The performance of the porous implant design was evaluated through implementation of a finite element bone remodelling algorithm and comparing the bone response with a femur with fully solid implant model, in terms of stress distribution and mass change. The results of the analysis showed improved bone formation on the bone-implant interface, and enhanced stress transmission to the surrounding bone from the implant.

# 1    Introduction

Join replacement surgeries, such as total hip replacement (THR), are one of the successful interventions among orthopaedic treatments, enhancing patient quality of life and relieving pain in the long-term. However, in some cases the orthopaedic implants fail, and a revision surgery is required. Aseptic loosening is one of the main causes of implant failures, whereby local bone resorption takes place at the bone-implant interface [1,2]. The vast majority of orthopaedic implants are made of solid metals, such as cobalt chromium alloys, 316L stainless steel, and titanium-based alloys. However, the stiffness of these materials are orders of magnitude higher than that of the surrounding bone [2–4]. The large stiffness mismatch at implant-bone interface results in stress shielding, whereby most of load is carried by implant instead of the bone tissue itself [4,5]. As bone is an adaptive tissue, the reduced mechanical loading on the bone results in local bone resorption and loss of bone mineral density in the region, which leads to contact loosening in the interface [2,6] and ultimately implant failure.

Several approaches have been used to address the issue of stress shielding, with the aim of reducing the stiffness of the hip implant stem. These approaches have targeted the geometric design of the stem region, material properties of the implant, or a combination of both geometrical and material changes [2,4]. Geometric design modifications include alterations to cross-section geometry design [7], addition of holes and ridges to the stem [8,9], and femoral stem length [10,11], although it remains challenging to resolve the stiffness mismatch or achieve optimum load distributions with macro-level changes to implant design. On the other hand, micro-level design modifications included applications of biomimetic polymer-composite materials [12]. With advances in additive manufacturing techniques, a novel category of materials has recently emerged that are referred to as meta-biomaterials, that is an intermediate concept that lies between material and structure [13]. The application of meta-materials in addressing stress shielding problem include using different functionally graded (FG) lattice structures in porous hip implant designs [5,14,15], and novel meta-biomaterials in orthopaedic devices design, such as auxetic and deployable meta-implants [1,16].

Porous lattice structures have been used both as surface coatings in implants to address issues with fixation by promoting bone ingrowth [2,17,18], and to address the stiffness mismatch between implant and bone using a more lightweight implant design [4,5,15,19–21]. This advantage has been clinically approved by evidence of enhanced bone regeneration using optimised 3D titanium-mesh scaffolds in sheep [22], with a computational optimisation framework developed accordingly [23]. However, many current hip implant designs with functionally graded structures are limited to uniform lattices [5,20], or simple gradient lattice structures [24–26]. Although, several more advanced lattice models have been suggested based on hybrid lattice designs [5,19], and optimised material distribution throughout the implant, using tetrahedron-based cells [3,4], and body centric cubic (BCC) structures [14,21]. The compatibility of neighbouring unit cells in functionally graded materials in hip implant design has been optimised through connectivity between adjacent microstructural unit cells in two-dimensions [27]. While these studies have introduced novel functionally graded implant structures, it is notable that many have not achieved true optimisation as their design process has not considered the long-term remodelling process of the bone [3,21,27], which is highly dependent on the dynamic loading conditions experienced at the bone-implant interface. Furthermore, certain approaches have not been able to map the density distribution to small sub-domains in a continuous manner [4]. Porous implant designs have also been proposed using triply periodic minimal surface (TPMS) structures [25,28], which have been mechanically investigated [29,30]. Such approaches have used simple uniform TPMS lattices [24,25], or multi-morphology lattice structures [31], representing graded mechanical requirements. However, while TPMS lattice structures offer morphological and mechanical properties that mimic trabecular bone, current approaches in hip implant design, have not yet sought to map optimised material distribution using TPMS.

The objective of this study is to develop a functionally graded biomimetic lattice-based hip implant design that is structurally optimised to limit stress shielding to the surrounding bone. The study develops and implements a new optimisation algorithm based on an inverse bone remodelling logic, whereby the optimised density distribution for the orthopaedic implants could be determined. Using the element-by-element density mapping lattice technique developed in LatticeWorks [32], a non-uniform functionally graded TPMS gyroid structure was mapped to the implant geometry, to produce the optimised structure, with a solid shell surrounding the porous structure. The long-term adaptive response of the bone tissue was evaluated using a strain energy density-based bone remodelling scheme, and the performance of the solid and functionally graded porous implant was compared. Also, this study includes an additively manufactured prototype of the implant, demonstrating its manufacturability.

## 2   Materials and methods

### 2.1   Model Geometry and Discretisation

A three-dimensional model was prepared using a hip implant and a femoral bone geometry. The implant model was created based on an identical configuration to a Zimmer M/L Taper implants (Zimmer Inc., USA, 2010). The bone was generated as a surface using a cloud point of a femur model from a previous study [33]. The generated parts with their dimensions are shown in Figure 1, with the design domain in hip stem region shown in green. To assemble the two parts, the implant stem surface was subtracted from the bone part, enabling the implant to be virtually inserted into the bone. The two parts, with linear elastic material properties (see section 3.1), were assembled as shown in Figure 1(c). Contact at the bone-implant interface was modelled as a node-to-surface interaction, with hard contact defined in the normal direction while the tangential direction was defined by a friction coefficient of 0.45 [34,35]. The two generated parts were meshed using linear tetrahedral elements (C3D4), using TetGen [36,37] algorithm, as shown in Figure 1(c). The FE model of solid implant and bone parts consisted of 3812 nodes and a total of 17,700 elements, and 13,295 nodes and 54,784 elements, respectively.

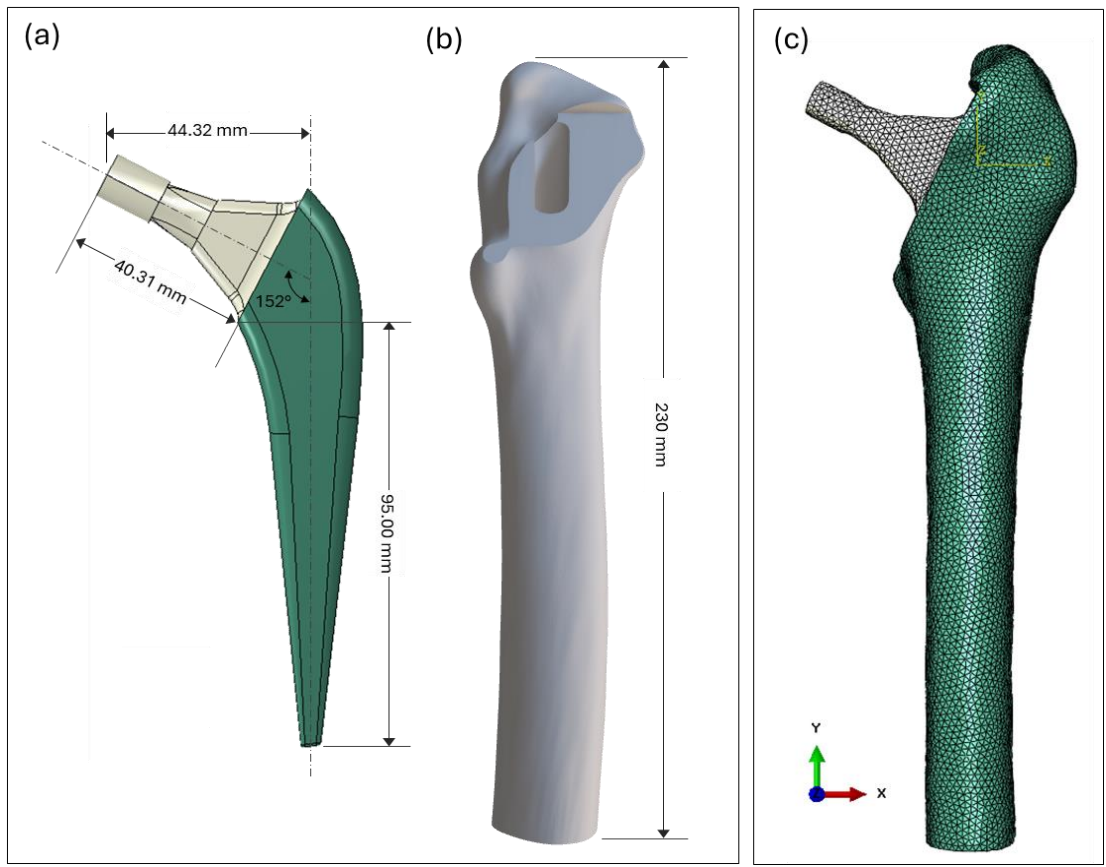

Figure 1 (a) Implant part with neck and stem regions, (b) femur cortical bone section, hollowed with the implant stem, (c) mesh intensity of implant and bone assembly.

## 2.2 Model Framework

### 2.2.1 General Overview

This study considered four distinct steps in developing a functionally graded porous implant. Firstly, an existing bone remodelling algorithm [38,39] was implemented on the bone section, to predict the bone tissue response to the implanted femoral implant and identify any instances of stress shielding. A second algorithm was developed to determine the optimized material distribution within the implant part, with the formulation of this algorithm based on the principle of an inverse remodelling logic. Subsequently, this optimised density distribution was mapped to a tailored density lattice structure on the implant through the appropriate mapping technique using the recently developed LatticeWorks toolbox [32]. LatticeWorks is an open-source MATLAB toolbox for lattice generation, facilitating creation of nonuniform and density mapping lattice structures, and infill lattice generation. A new implant design was suggested through mapping the optimised material distribution. Finally, the bone remodelling algorithm was applied once again on the femur with porous implant and the performance in terms of stress shielding evaluated and compared to the solid implant. The following sections describe the methods applied in each step in more detail.

*2.2.2  Bone Remodelling Algorithm*

A bone remodelling algorithm, inspired by Wolff's Law, where the strain energy density (SED) is used as a feedback control variable was implemented to determine the bone density distribution [38,39]. The algorithm adjusts density to balance SED around a reference value, triggering (i) positive remodelling (bone apposition), (ii) negative remodelling (bone resorption), or (iii) no change (lazy zone). The strain energy per unit of mass, $S$, was based on the physiological loads applied and was calculated as Equation (1).

$$S = \frac{1}{2\rho} . \sigma^T . \varepsilon \qquad (1)$$

in which $\rho$ represent the density, $\sigma$ is the tissue level stress based on element Von Mises stress, and $\varepsilon$ is the strain tensor from the FE analysis. The average strain energy density per unit of mass is calculated over the integration points of each element. The applied remodelling algorithm updated the material properties of elements, according to their average strain energy per unit of mass $S$, compared with a reference value, $S_{ref}$. The reference strain energy ($S_{ref}$) was calculated based on the initial state of bone, as the homeostatic stress and elastic modulus were $\sigma_{initial} = 68\ MPa$, and $E_{initial} = 17\ GPa$. The bone remodelling stimulus was the difference between the $S_{ref}$, and bone remodelling flag was defined as,

$$S - S_{ref} \begin{cases} > 0, & Positive\ Remodelling \\ = 0, & No\ Remodelling \\ < 0, & Negative\ Remodelling \end{cases} \qquad (2)$$

Based on the remodelling flag, the change an element's apparent density was calculated according to,

$$\Delta \rho = (S - S_{ref}) . \tau . A(\rho) . dt \qquad (3)$$

in which $dt$ is the time of the increment, $\tau$ is a time constant and assumed to be $\frac{130}{28} gmm^{-2} Jg^{-1}$ per day, and $A(\rho)$ is the free surface area, and a function of density, defined as [40],

$$A(\rho) = -0.8165\,\rho^6 + 2.7247\rho^5 - 4.7325\rho^4 + 5.8881\,\rho^3 - 7.6906\rho^2 \\ + 8.6563\rho + 0.1318 \tag{4}$$

The elements density is updated according to the remodelling flag, which was defined as,

$$\rho_{t+dt} = \begin{cases} \rho_t + |\Delta\rho|, & Positive\ Remodelling \\ \rho_t, & No\ Remodelling \\ \rho_t - |\Delta\rho|, & Negative\ Remodelling \end{cases} \tag{5}$$

The upper and lower density for the cortical bone elements was set to be $[0.01, 1.73]\ g.cm^{-3}$. When the density was calculated, the Young's modulus of an element was updated using Equation (6) [38,41].

$$E = 3790\,\rho^3 \tag{6}$$

Using the updated material properties at each increment, the FE solver updates the response of the system based on the boundary conditions. The maximum allowable density change in each iteration ($\Delta\rho_{max}$) was limited to $\frac{0.175}{28}\ g/cm^{-3}$ per day. The bone remodelling algorithm was considered converged when the change in density among the elements were lower than 0.002 in successive increments. The bone remodelling parameters are summarised in Table 1.

Table 1 Summary of bone remodelling parameters

| Remodelling parameter | value |
| --- | --- |
| **Homeostatic/Initial Stress ($\sigma_{initial}$)** | $68\ Mpa$ |
| **Remodelling time constant ($\tau$) [38]** | $\frac{130}{28} gmm^{-1}Jg^{-1}\ per\ day$ |
| **Max. cortical bone density [38]** | $1.73\ g.cm^{-3}$ |
| **Min. cortical bone density [38]** | $0.01\ g.cm^{-3}$ |
| **Max allowable change in density ($\Delta\rho_{max}$)** | $\frac{0.175}{28}\ g.cm^{-3}$ |

The bone remodelling algorithm is summarized in schematic flowchart shown in Figure 2, with the model implemented in the Abaqus/Standard [42] solver, through a user-defined field Subroutine (USDFLD).

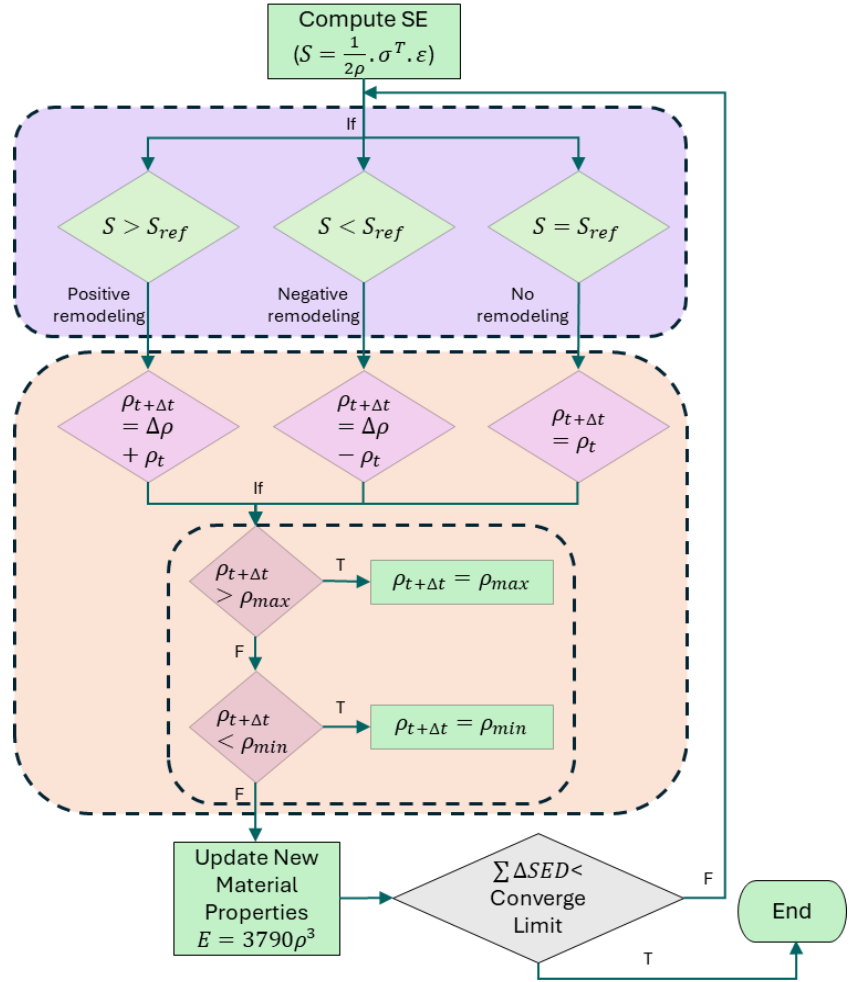

Figure 2 Bone remodelling algorithm, based on strain energy density, applied on the implanted femur section.

### 2.2.3 Density Optimisation Algorithm

An optimisation algorithm was developed based on an inverse-remodelling scheme to achieve a more homogenous stress distribution within an implant, by iteratively reducing stiffness via density adjustments in high-SED regions, promoting load redistribution.

The objective of the optimisation method was to reduce the total mass of the implant stem by 20%, while optimizing SED distribution. This topological objective, targeting material reduction in the final design, and was addressed by tuning the strain energy per unit of mass ($S$) of the elements, calculated using Equation (1), compared to a reference strain energy ($S_{ref}$). Mathematically, this is expressed as reducing the total mass ($M$) subject to the balanced SED ($S - S_{ref}$), where SED drives the optimization, aligning with bone remodelling principles. The optimisation logic is governed by Equation (7), in which the reference strain energy was considered $S_{ref} = 1e - 5 \, J/g$, based on the original strain energy of the elements to initiate the optimisation.

$$M = sum(\rho_i V_i), \quad \text{subject to: } S - S_{ref} \begin{cases} < 0, & \text{Increase stiffness} \\ = 0, & \text{No stiffness change} \\ > 0, & \text{Reduce stiffness} \end{cases} \quad (7)$$

in which $\rho_i$ and $V_i$ denote the apparent density and volume of the individual elements, respectively. The density change in each iteration was calculated as,

$$\Delta\rho = (S - S_{ref}).\tau.dt \quad (8)$$

In the optimisation process, the density change was not a function of the surface free area. Therefore, $A(\rho)$ factor was considered as 1 in this algorithm. Based on the strain energy difference and the stiffness optimisation scheme, the elements density was updated according to the remodelling flag,

$$\rho_{t+dt} = \begin{cases} \rho_t + |\Delta\rho|, & \text{Increase stiffness} \\ \rho_t, & \text{No stiffness change} \\ \rho_t - |\Delta\rho|, & \text{Reduce stiffness} \end{cases} \quad (9)$$

The lower and upper bounds for the implant element density were set between $[1.36, 4.5] \, g.cm^{-3}$, which corresponds to relative density of [30%-100%], using implant material properties [43]. When the density was calculated, an element's Young's modulus was updated based on the gyroid power law equation [29,30], using Equation (10).

$$E = 8120.2 \, \rho^{1.79} \quad (10)$$

The flowchart of the optimisation scheme is shown in Figure 3. While the optimization targets a 20% mass reduction, the ultimate goal is to mitigate stress shielding, assessed post-optimization via the bone remodelling algorithm (as per Section 2.2.2).

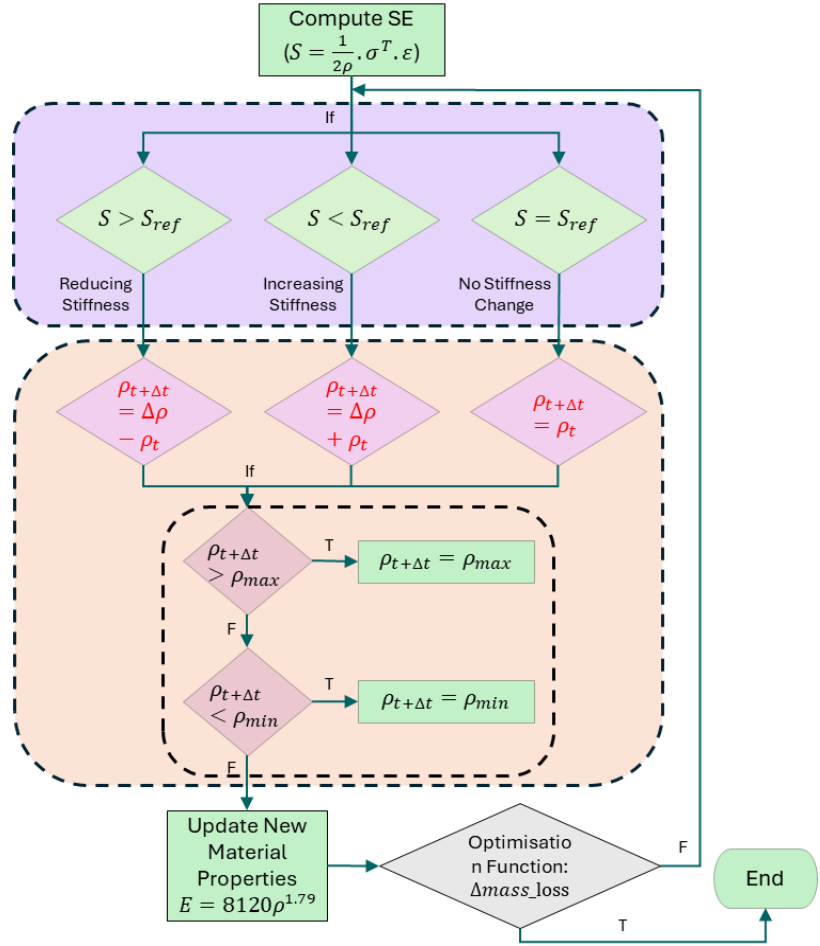

Figure 3 Stiffness optimisation algorithm, based on the inverse-remodelling algorithm, applied on the implant section.

*2.2.3.1    Functionally Graded Biomimetic Lattice Structure Tailoring Density Distribution*

The LatticeWorks toolbox [32], was used to map the optimised density distribution to a graded lattice structure. Using this toolbox, a grid was first generated in the bounding box enclosing the structure. An image matrix ($D$) was created using the grids, to calculate the distance of individual grids from the nearest boundary face of the implant. This distance matrix was binarized to inside ($D < 0$) and outside ($D > 0$) of the design region. The FE model density data was then interpolated to the inner grids, using barycentric coordinates of the elements. The grid relative density data was then translated to the gyroid levelset ($l$) data using the gyroid relative density vs levelset relationship and shown in Figure 4. The gyroid field function ($\varphi$) [29] was evaluated on the grid points ($X, Y, Z$), based on Equation (11), with the isotropic frequency of $70\pi$, that corresponded to the wavelength of $3.5\ cm$, in all three directions. Gyroid stut-based lattice choice with lower stiffness compared to sheet-based lattice [44] aligns with the goals of reducing stress shielding effect. The generated uniform gyroid field ($\varphi_{unifrom}$) was then normalised using the gradient levelset field ($l$), using Equation (12). The new gyroid field ($\varphi_{map}$) became non-uniform, mapping the graded density distribution.

$$\varphi_{uniform} = \sin X \cos Y + \sin Y \cos Z + \sin Z \cos X \quad (11)$$

$$\varphi_{map} = \frac{\varphi_{uniform}}{l} \quad (12)$$

Finally, the mapping gyroid field ($\varphi_{map}$) was rendered at isovalue of one, where the uniform gyroid field ($\varphi_{uniform}$) was equal to the levelset value ($l$), mapping the required relative density distribution.

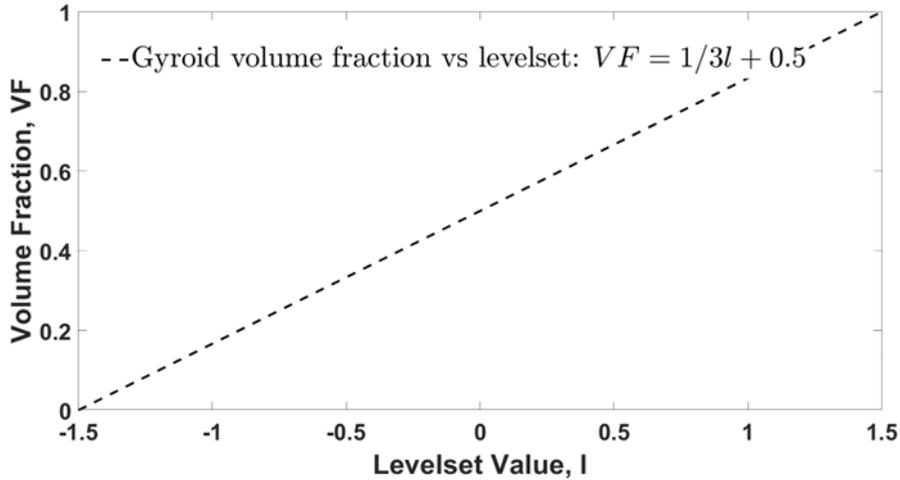

Figure 4 Gyroid relative density vs levelset relationship.

*2.2.3.2  Surface Shell Coating*

The generated gyroid field was enclosed in a shell with a thickness of $t = 4\ mm$, in terms of the distance value from the boundary surface of the implant stem in the image matrix. This step was done using density mapping with shell technique in LatticeWorks [32]. According to this technique, a copy of the distance matrix ($D$) was adopted to categorize the grids within a certain distance from the boundary face. The gyroid mapping field ($\varphi_{map}$) within the boundary region grids was replaced by the thickness field. With the new combined mapping and thickness field ($\varphi_{t,map}$), inner porous surface and boundary shells were drawn using isolevel value at zero.

## 3  Model Cases

*3.1  Bone Remodelling in Solid Implant*

With the FE model of the femur with fully solid implant prepared in Section 2.1, the bone remodelling algorithm was applied to the surrounding bone part. The material properties and the boundary conditions that were applied are described in the following sections.

In the first section for the analysis, the implant material was considered as titanium alloy, as an isotropic linear elastic material, with the mechanical properties reported in Table 2 [35,41]. The cortical bone section undergoes a remodelling algorithm, in which the material properties are updated in each solver increment. However, the allowed upper and lower limits of its Young's modulus in the algorithm and considered yield strength is reported in Table 2 [45,46].

Table 2 Mechanical Properties of human cortical bone tissue, and the implant.

| **Material Properties** | Elastic Modulus (GPa) | Poisson's Ratio | Yield Strength (MPa) |
|---|---|---|---|
| Implant (Ti-6AL-4V) | 110 | 0.3 | 900 |
| Cortical bone | 17 | 0.3 | Compression [130-200] Tension [50-151] |

The boundary conditions and applied loading for this case considered the static forces of a 700N person walking at normal speed [19,47,48]. The bone-implant assembly was subjected to the maximum muscle forces acting on the bone at node $P2$, and stem head contact force at node $P1$, as shown in Figure 5. A reduced muscle system was considered with a combination of abductors, the tensor fascia latae, and the vastus lateralis muscles, as summarised in Table 3. The bottom ends of the femoral bone were defined as fixed support, considering the contact surfaces of the knee joint. The applied boundary conditions are visualized in Figure 5.

Table 3 Loading components applied to the FE model.

| **Force Name (*Applied Location*)** | Force Components *(N)* | | |
|---|---|---|---|
| | $F_x$ | $F_y$ | $F_z$ |
| Hip joint (*P1*) | 378 | -1603 | 230 |
| Abductors (*P2*) | -406 | 605.5 | 30.1 |
| Vastus lateralis (*P2*) | 6.3 | -650.3 | 129.5 |
| Tensor fasciae latae distal part (*P2*) | 3.5 | -133 | 4.9 |
| Tensor fasciae latae proximal part (*P2*) | -50.4 | 92.4 | -81.2 |

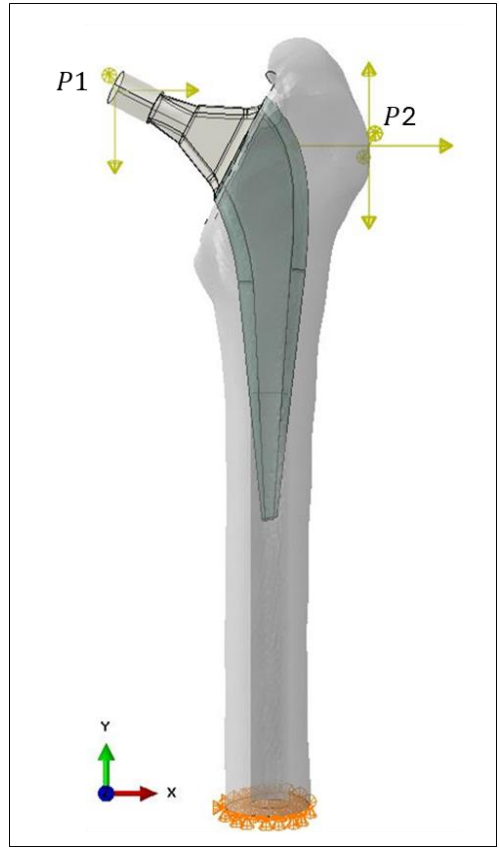

Figure 5 FE model with relevant physiological loadings and boundary conditions.

To evaluate the stress shielding effect in the bone-implant assembly, the FE model results were studied after convergence criteria was achieved, in 10 iterations of remodelling, and the bone mass was analysed on the bone elements. The femur bone was segmented into 7 Gruen zones [2,4], which are often used in clinical settings to evaluate THA performance. Also, to study the local effects of the implant, a bone-implant interface element set was defined on the bone section. These element groups are shown in Figure 6, for Gruen zones and bone-implant interface elements.

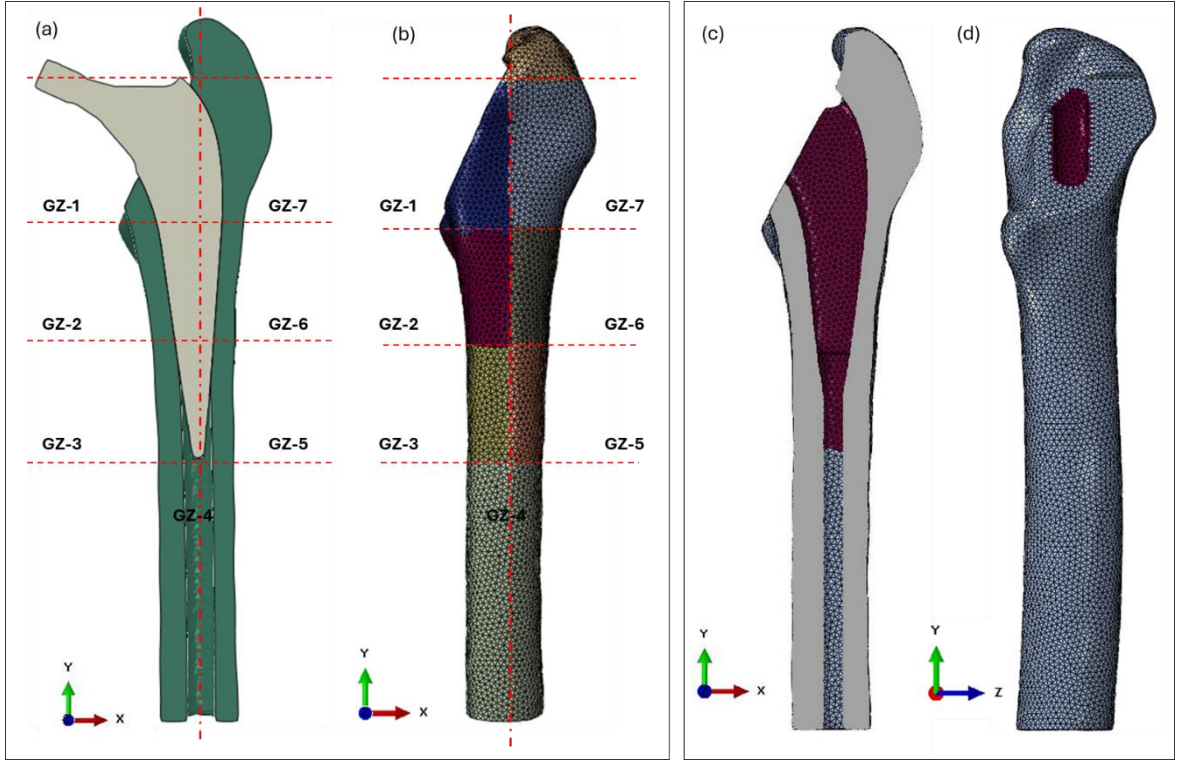

Figure 6 (a, b) Gruen zones, and (c, d) bone-implant interface element sets in red.

Bone element density, and element volume were exported from the results, and bone mass was calculated for the element sets defined. The change in elements bone mass was compared for the porous functionally graded implant design with the solid implant design, and the percentage of change in bone mass was reported based on,

$$\Delta m\ (\%) = \frac{m_{Porous\_implanted\_femur} - m_{Solid\_implanted\_femur}}{m_{Solid\_implanted\_femur}} \times 100 \tag{13}$$

### 3.2   Optimisation of Density Distribution and Functionally Graded Lattice Mapping

The optimisation algorithm was applied to the implant design section in the bone-implant assembly, as shown in Figure 1(c). In this assembly, the bone section was defined as an isotropic linear elastic material, with constant Young's modulus of $17\ GPa$, and Poisson ratio of 0.3. All boundary conditions and applied loads were the same as those described in bone remodelling model, as presented in Table 3, and Figure 5. The implant optimisation algorithm was considered converged when the mass reduction among the elements reached lower than 20% of the original mass. The optimisation algorithm was only applied to the design region of the implant, as shown in Figure 1(a), and the neck region maintained solid with material properties stated in Table 2.

### 3.3   Bone Remodelling in Functionally Graded Porous Implant

To evaluate stress shielding in the newly designed porous implant model, a FE model similar to the model in Section 2.1 was developed, with this implementation using the porous implant, instead of the solid implant. Following generation of the porous stem in LatticeWorks, it was merged with the neck region. The assembled implant was meshed using TetGen algorithm [36,37]. However, to mesh the solid section of the implant, hollowed gyroid voxels needed to be recognized by the meshing algorithm, otherwise it would automatically fill them up. To this end, the individual continuous holes were introduced to the algorithm by an inner point. However, due to resolution limitations, the mapped gyroid surfaces were not continuous, and there were groups of them throughout the stem region. To address this, continuous groups of gyroid surfaces were sorted based on their volumes. The largest regions, with a volume threshold of $500\ mm^3$ were selected, and any remaining sections or tiny gyroid surfaces were removed. With introducing an inner point of the maintained gyroid surfaces, the new implant model was meshed, and mesh data was converted to Abaqus input parameters. With the two FE models of implanted femur, one with a fully solid implant, and the other with the new porous design, the remodelling algorithm was applied to the surrounding bone part.

With the new implant design, the stress shielding effect was evaluated through bone mass in different element groups, through the method discussed in Section 3.1. The new implant design was assessed through comparing the results with the case of a femur with fully solid implant.

*3.4    Prototype Printing of Functionally Graded Porous Implant*

The resulting optimised functionally graded porous implant design was manufactured with Direct Metal Laser Sintering (DMLS) method, using aluminium, by LPE company (Laser Prototypes Europe Ltd., UK), using EOS M290 printer. DMLS is a very efficient sort of additive manufacturing procedure that uses a layer-by-layer manufacturing technique to produce any complex design. The main benefits of this technique are its adaptability to different materials and shapes, its vast potential for producing intricate 3D parts [49].

A high resolution stereolithography (STL) model of the designed implant was generated for 3D printing. The printing parameters are summarized in Table 4, in which for different printing directions, laser power and speed is specified.

Table 4 Printing process parameters.

|  | DownSkin | UpSkin | InFIll |
|---|---|---|---|
| **Laser Power (W)** | 210 | 370 | 360 |
| **Laser Speed (mm/s)** | 1600 | 1240 | 720 |
| **Hatch Distances (mm)** | 0.075 | 0.15 | 0.2 |
| **Layer Thickness (mm)** | 0.06 | | |
| **Powder Size (mm)** | 0.04 | | |

## 4 Results

### 4.1 Bone Remodelling in Solid Implant

Figure 7 shows the temporal evolution of Von Mises stress distribution, and subsequently bone density, in the femur following implantation of the bone remodelling algorithm on fully solid implant design. Stress values at the outer layers of cortical region triggered a positive remodelling scheme, and the bone density increased in these regions. As shown in Figure 7(b), the density distribution started from a homogenous state of $\rho_{initial} = 1.65 \ g/cm^3$, which is notable that this state was not a homeostatic condition. This means that the bone remodelling algorithm showed distinct changes in the early iterations, with high bone density remaining in high stress regions and regions of bone resorption detected in non-load carrying regions in the proximal femur.

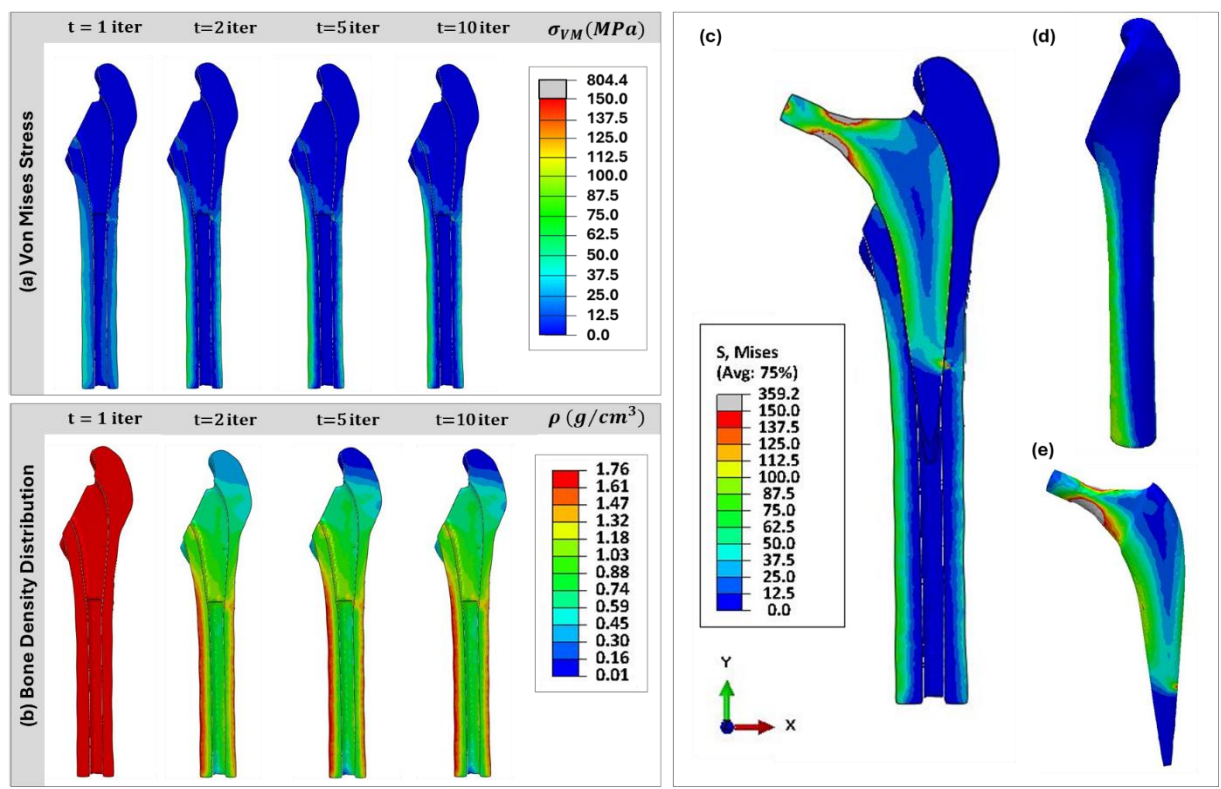

Figure 7 Bone remodelling results on femur with fully solid implant. (a) Development of Von Mises stress distribution, and (b) bone density variation over time, for femur with fully solid implant, under remodelling algorithm. Final Von Mises stress distribution within (c) cut view of femur with fully solid implant assembly, (d) bone part and (e) solid implant part.

Figure 7(c) shows the Von Mises stress distribution in the assembly of a femur with fully solid implant, following the bone remodelling algorithm is converged. Figure 7(c) shows that at the bone-implant interface, the solid implant elements are under higher stress levels, compared to the surrounding bone. In the bone section in Figure 7(d), the circumferential cortical region carries more stress compared to the concentric bone elements within bone canal, and stress reduces as getting closer to near-the-implant trabecular region. The implant section on the other hand in Figure 7(e), shows an uneven stress distribution throughout the stem region, with concentrated stress in the bone-implant interface. Maximum stress in the implant was $359\ MPa$, in the neck region of the implant.

*4.2  Optimisation of Density Distribution and Functionally Graded Lattice Mapping*

Figure 8 shows how the bone density distribution and Von Mises stress evolved over time, as a result of the optimisation algorithm. Here, the element density was originally $4.06\ g/cm^3$ throughout the entire structure. During the optimisation process, the density and stiffness started to reduce on the sides of the proximal region of the stem. Low density regions evolved horizontally and vertically over time. On the other hand, the density in the distal region did not reduce over the analysis time. The final density distribution is shown at time $t = 6.16$, where total mass reduction within the structure reached 20%.

Figure 9(a, b) shows the resulting non-uniform density distribution in the design region of the implant, as the result of the optimisation algorithm on the solid implant. In this suggested material distribution, the core of the implant had greater density, whereas the density of material was reduced on the sides of the medial axis in the implant. The reduced density regions were under higher stresses, as shown in Figure 8(b) and went under the *reducing stiffness* loop in the inverse remodelling algorithm. A minimum allowable density of $1.36\ g/cm^3$ corresponds to the volume fraction of 30% in elements. The narrow bottom part of the stem also maintained higher density values, as it was not under high stress values. The density distribution was translated to a relative density distribution, and according to the gyroid power law equations and using Equation (10), and Young's modulus distribution map over the design region was calculated, as shown in Figure 9(b).

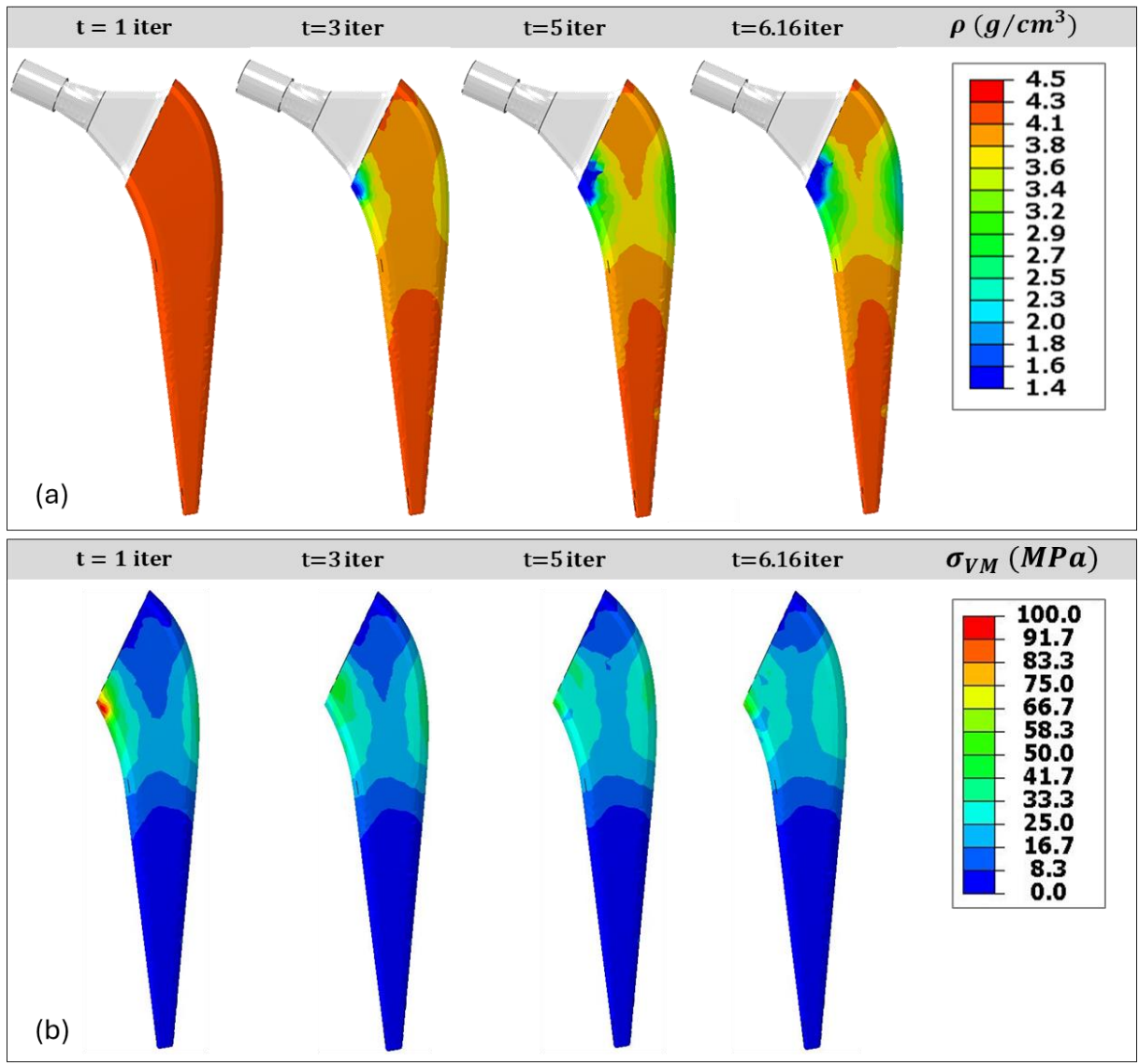

Figure 8 (a) Density distribution, and (b) Von Mises stress in the design region of implant as a function of time.

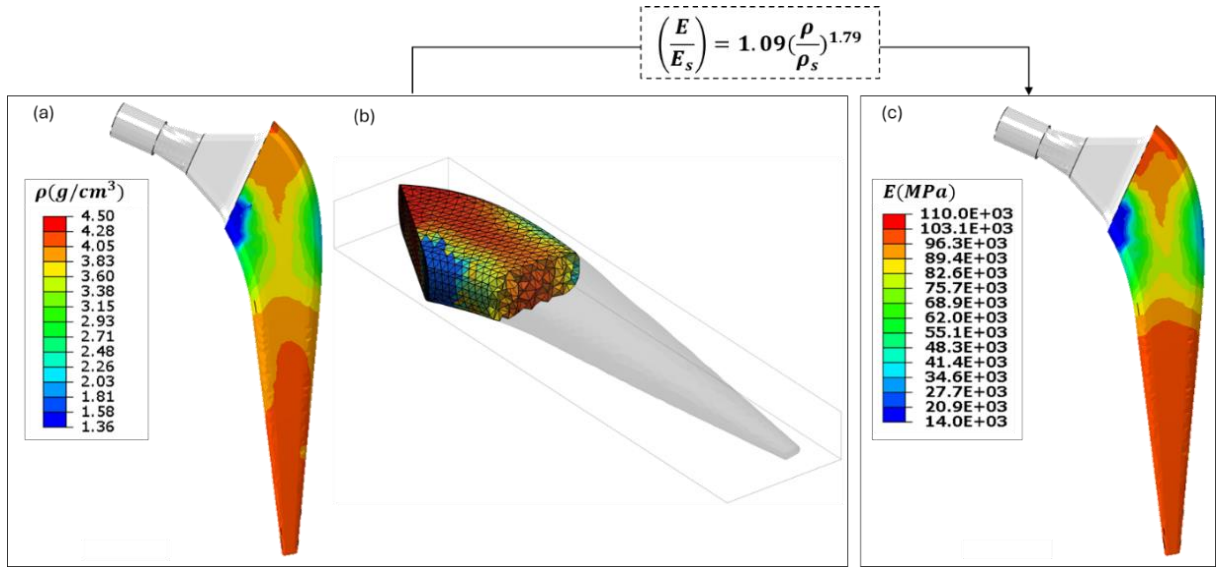

Figure 9 (a) Optimised density distribution in the design region of the implant, (b) a cross section of the density distribution, (c) corresponding elastic modulus distribution on the design region.

*4.2.1 Functionally Graded Lattice Mapping*

The results of the mapping techniques are shown as step-by-step in Figure 10. The binarized grid, shown in Figure 10(a), divides the domain into points that are inside the implant surface and the outside points. The interpolated element density data, as the results of FE optimisation, calculated for the inner grids is shown in Figure 10(b). Based on the relative density vs levelset equation in gyroid structure in Figure 4, the gyroid levelset distribution field was calculated, shown in Figure 10(c), and the density distribution field was mapped into a gyroid levelset field. The equivalent gyroid levelset variation was calculated to be [-1.5, 0.75], for volume fraction variation of [0.3, 1]. The normalized mapped gyroid field by the levelset values, was rendered at isovalue of one and is shown in Figure 10(d). This new gyroid field is non-uniformly graded, and captures the same gradient map, as shown in density distribution field in Figure 9, enabling to draw the gyroid isosurface and create the porous structure.

The resulting graded gyroid lattice surface with optimised density distribution is shown in Figure 10(e). The gyroid surface, shown in grey in Figure 10(e), represents a porous region that was needed to be removed, to capture the derived density distribution map within the implant region. The other surface is the implant original boundary surface which is shown in pink in Figure 10(e). To construct the implant, the area between the two surfaces was solidified, that the porous gyroid was deducted from it.

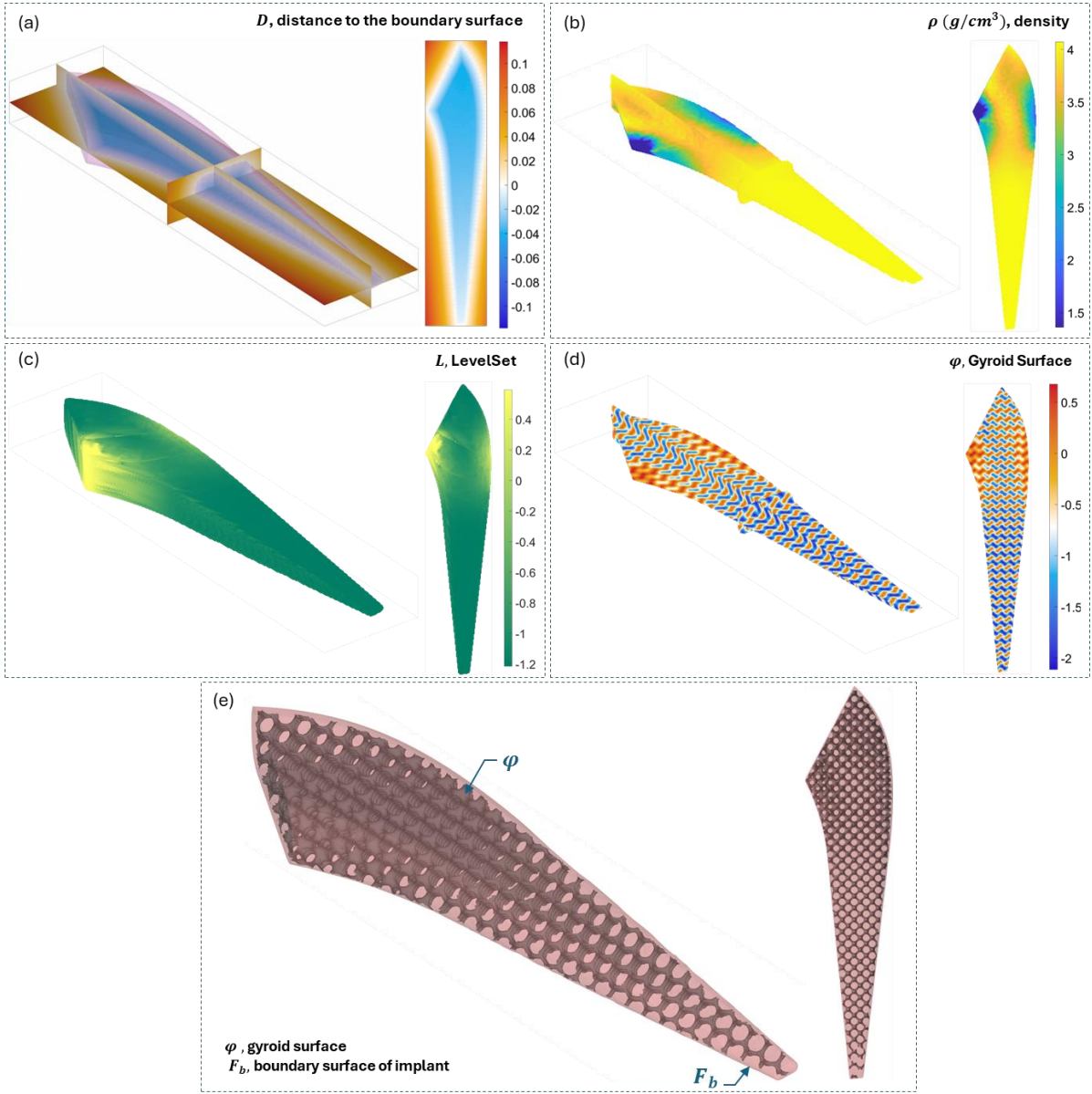

Figure 10 Mapping Density lattice generation, (a) binarized grid over the bounding box of the implant stem, (b) element density distribution interpolated on the grid, (c) equivalent gyroid levelset field to the density distribution, (d) gyroid function values created on the levelset field, (e) isosurfaces on the generated mapping gyroid field in grey, and outer face boundary surface of the implant shown in pink.

### 4.2.2 Surface Shell Coating

The generated the field for the coating shell ($R$) is shown in Figure 11(a), in which the coating shell is shown as negative field values. This shell field overwrites the gyroid field function ($\varphi$) on the surface boundary with thickness, $t$. The combined field function of these two fields were rendered at isovalue of one, and the resultant surfaces is shown in Figure 11(b). The inner grey surface is generated based on the gyroid field function, and the surrounding pink surface is generated based on the original implant boundary surface, and the thickness gap is applied in between these two surfaces. With the gyroid voids surface inside, and a cover surface on the boundaries, the space between the two surfaces was solidified, and slice images of the final design is shown in Figure 11(c).

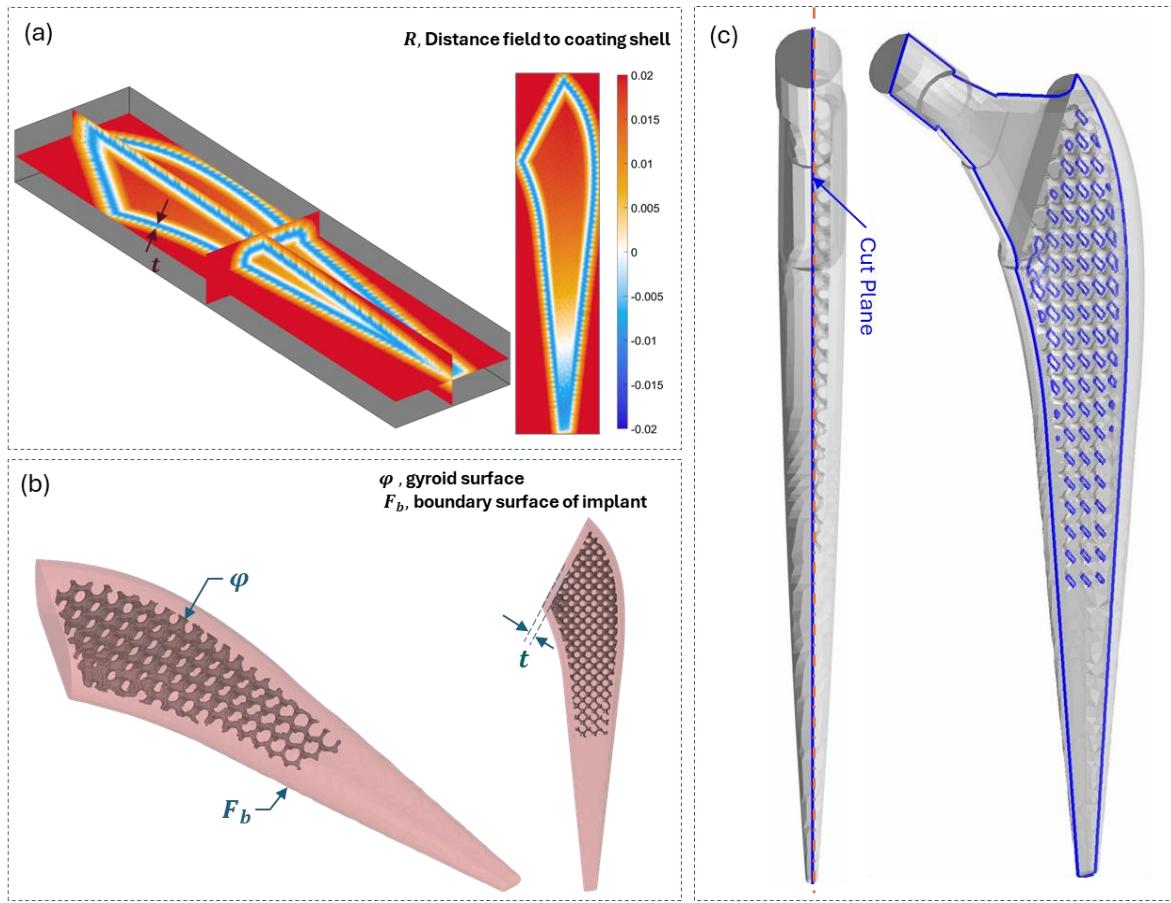

Figure 11 Shell coating, (a) shell thickness t, on the boundary of the implant stem surface, (b) gyroid surface (in grey), and coating surface (in pink) for generating the porous implant, (c) slicers of the designed porous implant.

*4.3    Bone Remodelling in Functionally Graded Porous Implant*

The results of the bone remodelling algorithm following implantation of the functionally graded porous implant were compared to the femur with fully solid implant. Figure 12 shows density and elastic modulus of the surrounding bone, for the femur with porous and fully solid implant, following the bone remodelling algorithm. According to Figure 12(a), the density of elements was higher in the bone-implant interface, when implanted with the functionally graded porous design. As a result, the elastic modulus of the elements was higher in these regions, as shown in Figure 12(b), for the porous implant case compared to the solid implant.

The bone mass of element sets in the two cases of femur with fully solid and porous implant is summarised in Figure 12(c). The percentage of change in bone mass with the porous design implant, compared to the solid implant design, is reported for Gruen zones and bone-implant interface elements, as shown in Figure 12(d, e). According to these results, the bone mass particularly increased in Gruen zones 3 and 5. However, Gruen zones 6 and 7 showed mass reduction, and Gruen zones 1, 2 and 4 did not show any considerable mass change, comparing the porous implant with the solid implant design. Although, the total bone mass with the two implant cases were almost similar, the local changes of bone mass on the bone-implant interface elements, showed an overall mass increase of 33%, based on Figure 12(e). According to the results, reducing the implant elastic modulus created a local increase in bone mass for the surrounding bone, although not changing the overall bone mass.

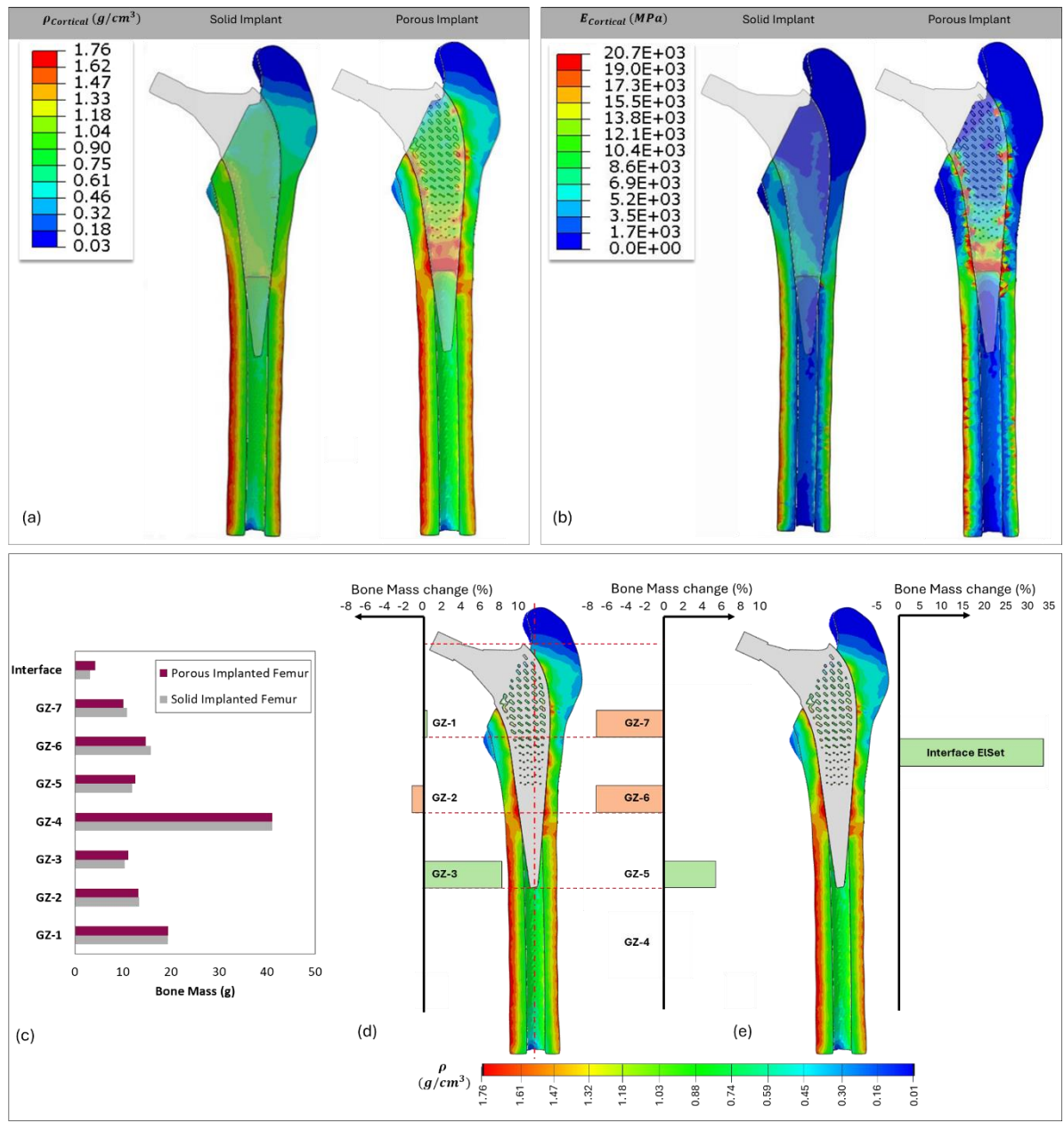

Figure 12 Bone remodelling results on femur with fully solid implant vs porous implant. (a) Density and (b) elastic modulus distribution on the femur with fully solid vs porous. (c) Bone mass results in femur with fully solid vs porous implant, categorized by (d) Gruen zones, and (e) interface bone elements.

Figure 13 shows the Von Mises stress distribution on the bone, implants, and assembled systems for both implant designs. In the case of the functionally graded porous implant, as shown in Figure 13(a, b), bone elements experienced higher stress at the bone-implant interface region. Subsequently, in the high stress regions, positive remodelling was triggered in those elements, and higher density was reported, as shown in Figure 12(a).

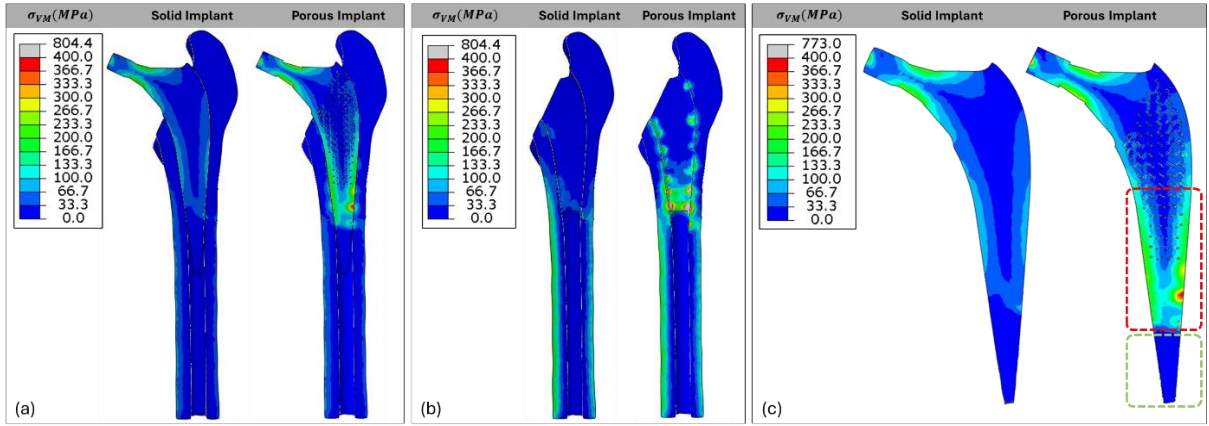

Figure 13 Von Mises stress distribution in the porous-implant compared to the solid-implanted femur (a) assembly, (b) bone parts, and (c) implant parts.

According to Figure 13(c), in the functionally graded porous implant design, the stress has increased on the solid shell region, to compensate for the material reduction. The stress distribution has also been shifted more towards the medial axes of the implant, in the porous implant design. The middle section of the functionally graded porous implant, shown in red window in Figure 13(c), had lower porosity in and was also under higher stress levels compared to the same regions in the fully solid implant design. However, in the lower parts of the functionally graded porous implant, shown in green window in Figure 13(c), elements were not in direct surface contact with the surrounding bone. Therefore, they showed lower stress values. There is a small region on the porous implant with high stress concentration, which is due to the contact with the geometrical artifacts in bone marrow. Moreover, the neck region of the functionally graded porous implant, as the critical region, still maintained a safety factor of 2 ($SF = \frac{\sigma_{yield}}{\sigma_{max}}$) [2], considering the yield stress of the material.

*4.4 Prototype Printing of Functionally Graded Porous Implant*

To demonstrate the internal porous structure of the implant design, it was printed in two sections, cut from the medial plane, as shown in Figure 14(a), Figure 14(b) shows the printed prototype. Due to the fact that components manufactured by DMLS have typically as-fabricated surface defects [50], the inner pore surfaces include particles of un-melted metal powder or have pores created as a result of weld pool collapses. However, the resolution of the model, the minimum pore size in the model, and particle size affect the final product quality.

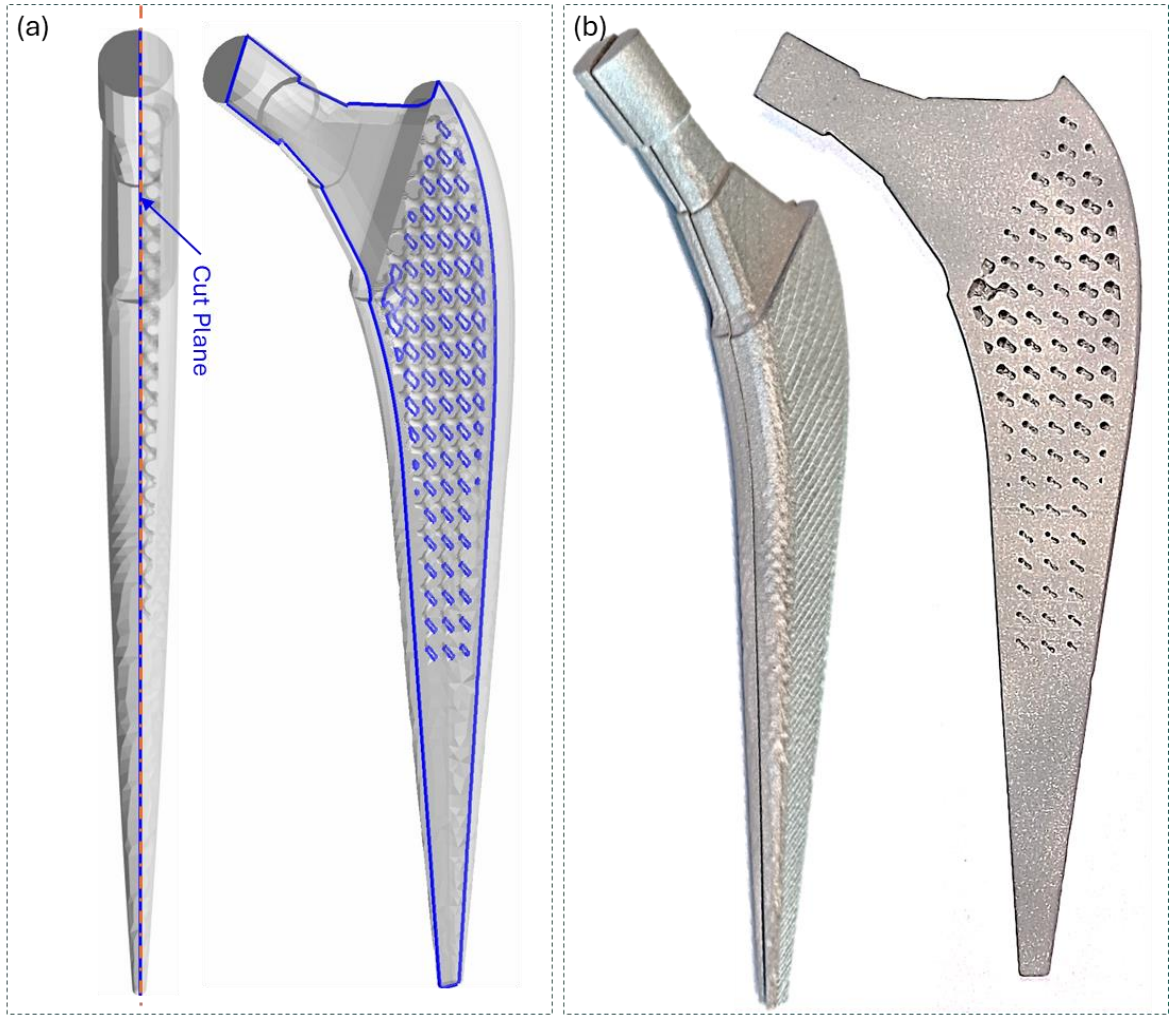

Figure 14 (a) An implant specimen model and its cross-section, (b) fabricated using direct metal laser sintering (DMLS). Material: Aluminium, with particle size of 40 μm. Machine: EOS M290, LPE Company, Belfast, UK.

## 5   Discussion

This study developed a functionally graded hip implant design using a gyroid lattice-based structure that was structurally optimised through an inverse bone remodelling algorithm. The functionally graded gyroid lattice structure was mapped on an optimised density distribution field in the implant stem region. The density optimisation technique was based on an inverse bone remodelling logic, which produced a structure that had low density regions on the sides of the medial axis of the stem and in the proximal region, while having a higher density at distal part of the stem (Figure 9). The performance of the functionally graded porous implant design was evaluated using a bone remodelling algorithm. Local increases in the bone mass distribution were found for the functionally graded porous design, with higher stress successfully transferred to the bone elements at the bone-implant interface following implantation.

The optimised porous implant design that was developed in this study represents a novel design, with notable differences to existing commercially available porous implant designs that have not been optimised based on loading distribution. For example, the Arcos one-piece femoral revision system (Zimmer Biomet, USA, 2016) simply uses a randomly distributed porous structures throughout the component. Other devices tend to only include porosity on surface [51], as shown in Figure 15. Considering new additively manufactured designs of functionally graded porous implants, very few of them are based on an optimised material distribution [3,4,52], and in most cases, the gradient structure is a uniform or simple linear graded lattice structure [25,53], or else use hybrid designs, including multi-material lattice structures [1,5,19,54]. However, the graded lattice pattern that was produced in the current study is non-uniform and dictated by a material optimisation algorithm and it cannot simply be described by a linear gradient lattice. Here, the density mapping technique was used in the graded lattice design, that follows an element-by-element tailored structural properties, instead of using multi-morphology lattice design.

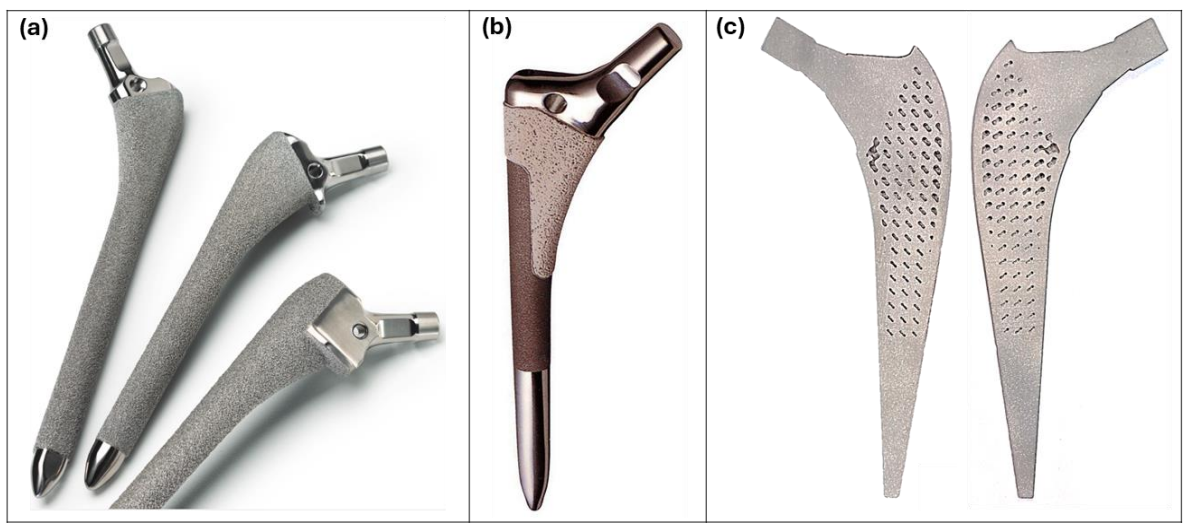

Figure 15 Commercially available (a) fully porous Arcos one-piece femoral revision system implant [from [55]] and (b) VerSys HA/TCP Fiber Metal Taper stem porous coated implant [from [51]], vs (c) the new graded porous implant design.

While the resulting distribution map was aligned with other studies that have used alternative optimisation algorithms [3,4], the element-by-element density mapping approach outlined here enabled a higher resolution grid point, which locally mapped continuous porous structure. This is in contrast to other methods [4,15] that limited the number of sampling points to reduce computational cost, with these distributed through the structure to map the density distribution. In addition, this study provides the option to cover the porous implant using a solid shell, to offer the opportunity for applying an optimized porous coating on the surface, whereas in most gradient porous implant designs, the body of implant consists of lattice structures. This generated a porous implant with gradient cell sizes and porosity on its surface [4,14,27]. Although the porosity might address the fixation by promoting bone growth in these implant designs, it needs more invasive techniques in case of a revision surgery [56,57]. The coated design was applied on the graded porous implant design, whereas in another study it was considered on a uniform BCC lattice structure [53]. Density-mapping lattices used in other studies were strut-based lattices including BC lattices [58], or tetrahedron lattices [3,4]. Although TPMS-based lattices have been used in hybrid lattice-based implant designs [5], however as a density-based lattice structure they have not been investigated before. Using the density mapping technique, introduced in LatticeWorks [32], the results of the optimisation section were used to generate a functionally graded lattice structure in the implant surface, that had the optimised material distribution mapped.

The new implant design was evaluated and compared to the fully solid implant design, using a bone remodelling algorithm on the surrounding bone. The quantified stress levels and bone density showed locally higher stress and higher density on the surrounding bone elements, in the case of the porous implant design. The increased interface stress was an undesirable effect of the new design, which might increase the probability of failure in bone [4,15]. However, in this study, the resulting stresses in the optimised implant upon loading were safe when evaluated and compared to the yield strength of the material. Also, the increased stress value in the bone tissue served as a positive remodelling stimulus for the interface elements. Bone mass results showed a 33% increase on the bone-implant interface elements in the femur with porous implant, compared to the femur with fully solid implant. Although, the results did not show considerable bone mass increase in the wider bone regions. The high stress regions in the implant and bone assembly were due to the geometry artifact and the step generated in the bone canal, while excavating the implant hole. This study also has several limitations that should be considered. After THA, a patient's most common dynamic activity is walking. Therefore, to calculate the changes in the physiological strain distribution following THA or bone remodelling, researchers typically looked at static load situations of the gait cycle. [41]. However, in normal life, femur is under other loading conditions as well, that incorporates all the muscle loads. Therefore, the design can be evaluated in more loading conditions, to capture a more inclusive bone response, that incorporates the upper region of the femur bone as well, whereas in this design was shown to be under resorption and reduced in density. Also, stress shielding effect can be quantified using stress shielding increase (SSI) indicator [2], that evaluates the response of the implanted femur with intact femur response. For further assessment of stress shielding effect, SSI factor can also be calculated to compare the two solid and porous implant design. However, in this work, the change of bone mass has been reported to assess the performance of the new implant, with reference to the femur with solid implant. Furthermore, the sectioned prototype of the implant design was also 3D printed to demonstrate manufacturability of the design, as well as additive manufacturing challenges. In this prototype, the sample was printed as two sides of the whole model. However, with the full model design, using powder bed technology, removing the remaining powder within the closed porous surface is not accessible as it is in the cut model. The additive manufacturing experience in this study showed that for future similar designs, a vent or powder removal channel needs to be considered at design stage, to facilitate prototyping with DMLS technique. Also, aluminium is not biocompatible within human body, and the real implants are available in other biocompatible materials [59–61]. In this demonstration of metal additive manufacturing, the material itself was not a matter of study. However, in future studies, it is recommended that the implant be printed in biocompatible materials. With the full implant model additively manufactures, the mechanical performance of the implant can be experimentally tested in terms of yield strength or fatigue in future works. However, this study is an inclusive package from modelling, optimisation, analysis, and prototyping that establishes the key aspects of the functionally graded porous implant design. The process can be modified and customised in each step according to the specific application.

# 6 Conclusion

This study developed an optimised functionally graded porous implant design, equipped with a solid shell coating, that showed improved long-term bone response. The optimisation technique was inspired by an inverse bone remodelling logic, that promoted an even stress distribution throughout the design region, by reducing the density and consequently the stiffness in regions where strain energy was higher than the reference level. The results of the optimisation technique were a density distribution field, that showed lower density on the sides of the implant stem, and higher material density around the medial axis, and distal regions of the implant. The performance of the optimised functionally graded porous implant design showed improved bone formation on the bone-implant interface. This study demonstrates how the computational frameworks developed in this work can be utilised to create functionally graded lattice structures, incorporating structural optimization techniques to facilitate customised implant design.

# 7 Acknowledgment

This project has received funding from the European Research Council (ERC) under the EU's Horizon 2020 research and innovation program (Grant agreement No. 804108).

# 8 Data Availability

The datasets generated and/or analysed during the current study are available in the Zenodo repository, https://zenodo.org/records/14906052.

# 9 References

[1] M.J. Mirzaali, A.A. Zadpoor, Orthopedic meta-implants, 010901 (2024). https://doi.org/10.1063/5.0179908.

[2] S.A. Naghavi, C. Lin, C. Sun, M. Tamaddon, M. Basiouny, P. Garcia-Souto, S. Taylor, J. Hua, D. Li, L. Wang, C. Liu, Stress Shielding and Bone Resorption of Press-Fit Polyether–Ether–Ketone (PEEK) Hip Prosthesis: A Sawbone Model Study, Polymers (Basel) 14 (2022). https://doi.org/10.3390/polym14214600.

[3] Y. Wang, S. Arabnejad, M. Tanzer, D. Pasini, Hip Implant Design With Three-Dimensional Porous Architecture of Optimized Graded Density, Journal of Mechanical Design 140 (2018). https://doi.org/10.1115/1.4041208.

[4] S. Arabnejad, B. Johnston, M. Tanzer, D. Pasini, Fully porous 3D printed titanium femoral stem to reduce stress-shielding following total hip arthroplasty, Journal of Orthopaedic Research 35 (2017) 1774–1783. https://doi.org/10.1002/jor.23445.

[5] S.A. Naghavi, M. Tamaddon, P. Garcia-Souto, M. Moazen, S. Taylor, J. Hua, C. Liu, A novel hybrid design and modelling of a customised graded Ti-6Al-4V porous hip implant to reduce stress-shielding: An experimental and numerical analysis, Front Bioeng Biotechnol 11 (2023) 1–20. https://doi.org/10.3389/fbioe.2023.1092361.

[6] P.T. Avval, S. Samiezadeh, V. Clav Klika, H. Bougherara, Investigating stress shielding spanned by biomimetic polymer-composite vs. metallic hip stem: A computational study using mechano-biochemical model, (2014). https://doi.org/10.1016/j.jmbbm.2014.09.019.

[7] A.L. Sabatini, T. Goswami, Hip implants VII: Finite element analysis and optimization of cross-sections, Mater Des 29 (2008) 1438–1446. https://doi.org/10.1016/j.matdes.2007.09.002.

[8] M. Ceddia, B. Trentadue, Evaluation of Rotational Stability and Stress Shielding of a Stem Optimized for Hip Replacements—A Finite Element Study, Prosthesis 5 (2023) 678–693. https://doi.org/10.3390/prosthesis5030048.

[9] S. Gross, E.W. Abel, A finite element analysis of hollow stemmed hip prostheses as a means of reducing stress shielding of the femur, J Biomech 34 (2001) 995–1003. https://doi.org/10.1016/S0021-9290(01)00072-0.

[10] B. Van Rietbergen, R. Huiskes, Load transfer and stress shielding of the hydroxyapatite-ABG hip: A study of stem length and proximal fixation, J Arthroplasty 16 (2001) 55–63. https://doi.org/10.1054/ARTH.2001.28369.

[11] A. Sas, P. Pellikaan, S. Kolk, P. Marty, T. Scheerlinck, G.H. van Lenthe, Effect of anatomical variability on stress-shielding induced by short calcar-guided stems: Automated finite element analysis of 90 femora, Journal of Orthopaedic Research 37 (2019) 681–688. https://doi.org/10.1002/JOR.24240.

[12] H. Bougherara, M. Bureau, M. Campbell, A. Vadean, L. Yahia, Design of a biomimetic polymer-composite hip prosthesis, J Biomed Mater Res A 82A (2007) 27–40. https://doi.org/10.1002/JBM.A.31146.

[13] S.A. Yavari, S.M. Ahmadi, R. Wauthle, B. Pouran, J. Schrooten, H. Weinans, A.A. Zadpoor, Relationship between unit cell type and porosity and the fatigue behavior of selective laser melted meta-biomaterials, J Mech Behav Biomed Mater 43 (2015) 91–

100. https://doi.org/10.1016/J.JMBBM.2014.12.015.

[14] C. Sun, J. Kang, L. Wang, Z. Jin, C. Liu, D. Li, Stress-dependent design and optimization methodology of gradient porous implant and application in femoral stem, Comput Methods Biomech Biomed Engin 0 (2022) 1–12. https://doi.org/10.1080/10255842.2022.2115291.

[15] S. Arabnejad Khanoki, D. Pasini, Multiscale Design and Multiobjective Optimization of Orthopedic Hip Implants with Functionally Graded Cellular Material, J Biomech Eng 134 (2012). https://doi.org/10.1115/1.4006115.

[16] H.M.A. Kolken, C.P. de Jonge, T. van der Sloten, A.F. Garcia, B. Pouran, K. Willemsen, H. Weinans, A.A. Zadpoor, Additively manufactured space-filling meta-implants, Acta Biomater 125 (2021) 345–357. https://doi.org/10.1016/J.ACTBIO.2021.02.020.

[17] P. Heinl, L. Müller, C. Körner, R.F. Singer, F.A. Müller, Cellular Ti–6Al–4V structures with interconnected macro porosity for bone implants fabricated by selective electron beam melting, Acta Biomater 4 (2008) 1536–1544. https://doi.org/10.1016/J.ACTBIO.2008.03.013.

[18] J. Lv, P. Xiu, J. Tan, Z. Jia, H. Cai, Z. Liu, Enhanced angiogenesis and osteogenesis in critical bone defects by the controlled release of BMP-2 and VEGF: Implantation of electron beam melting-fabricated porous Ti6Al4V scaffolds incorporating growth factor-doped fibrin glue, Biomedical Materials (Bristol) 10 (2015). https://doi.org/10.1088/1748-6041/10/3/035013.

[19] M.G. Gok, Creation and finite-element analysis of multi-lattice structure design in hip stem implant to reduce the stress-shielding effect, Proceedings of the Institution of Mechanical Engineers, Part L: Journal of Materials: Design and Applications 236 (2022) 429–439. https://doi.org/10.1177/14644207211046200.

[20] H.E. Burton, N.M. Eisenstein, B.M. Lawless, P. Jamshidi, M.A. Segarra, O. Addison, D.E.T. Shepherd, M.M. Attallah, L.M. Grover, S.C. Cox, The design of additively manufactured lattices to increase the functionality of medical implants, Materials Science and Engineering: C 94 (2019) 901–908. https://doi.org/10.1016/J.MSEC.2018.10.052.

[21] Y. He, D. Burkhalter, D. Durocher, J.M. Gilbert, Solid-Lattice Hip Prosthesis Design: Applying Topology and Lattice Optimization to Reduce Stress Shielding From Hip Implants, (2018) 1–5. https://doi.org/10.1115/dmd2018-6804.

[22] A.M. Pobloth, S. Checa, H. Razi, A. Petersen, J.C. Weaver, K. Chmidt-Bleek, M. Windolf, A.A. Tatai, C.P. Roth, K.D. Schaser, G.N. Duda, P. Schwabe, Mechanobiologically optimized 3D titanium-mesh scaffolds enhance bone regeneration in critical segmental defects in sheep, Sci Transl Med 10 (2018). https://doi.org/10.1126/scitranslmed.aam8828.

[23] C. Perier-Metz, G.N. Duda, S. Checa, A mechanobiological computer optimization framework to design scaffolds to enhance bone regeneration, Front Bioeng Biotechnol 10 (2022) 980727. https://doi.org/10.3389/FBIOE.2022.980727/FULL.

[24] E. Davoodi, H. Montazerian, R. Esmaeilizadeh, A.C. Darabi, A. Rashidi, J. Kadkhodapour, H. Jahed, M. Hoorfar, A.S. Milani, P.S. Weiss, A. Khademhosseini, E. Toyserkani, Additively Manufactured Gradient Porous Ti−6Al−4V Hip Replacement

Implants Embedded with Cell-Laden Gelatin Methacryloyl Hydrogels, Cite This: ACS Appl. Mater. Interfaces 13 (2021) 22110–22123. https://doi.org/10.1021/acsami.0c20751.

[25] N. Kladovasilakis, K. Tsongas, D. Tzetzis, Finite Element Analysis of Orthopedic Hip Implant with Functionally Graded Bioinspired Lattice Structures, Biomimetics 2020, Vol. 5, Page 44 5 (2020) 44. https://doi.org/10.3390/BIOMIMETICS5030044.

[26] A. Seharing, A. Hadi Azman, S. Abdullah, Finite element analysis of gradient lattice structure patterns for bone implant design, International Journal of Structural Integrity 11 (2020) 535–545. https://doi.org/10.1108/IJSI-03-2020-0028.

[27] E. Garner, H.M.A. Kolken, C.C.L. Wang, A.A. Zadpoor, J. Wu, Compatibility in microstructural optimization for additive manufacturing, Addit Manuf 26 (2019) 65–75. https://doi.org/10.1016/J.ADDMA.2018.12.007.

[28] K.G. Mostafa, G.A. Momesso, X. Li, D.S. Nobes, A.J. Qureshi, Dual Graded Lattice Structures: Generation Framework and Mechanical Properties Characterization, Polymers 2021, Vol. 13, Page 1528 13 (2021) 1528. https://doi.org/10.3390/POLYM13091528.

[29] M. Vafaeefar, K.M. Moerman, M. Kavousi, T.J. Vaughan, A morphological, topological and mechanical investigation of gyroid, spinodoid and dual-lattice algorithms as structural models of trabecular bone, J Mech Behav Biomed Mater 138 (2023) 105584. https://doi.org/10.1016/J.JMBBM.2022.105584.

[30] M. Vafaeefar, K.M. Moerman, T.J. Vaughan, Experimental and computational analysis of energy absorption characteristics of three biomimetic lattice structures under compression, J Mech Behav Biomed Mater 151 (2024) 106328. https://doi.org/10.1016/J.JMBBM.2023.106328.

[31] S. Zeng, G. Liu, W. He, J. Wang, J. Ye, C. Sun, Design and performance prediction of selective laser melted porous structure for femoral stem, Mater Today Commun 34 (2023) 104987. https://doi.org/10.1016/J.MTCOMM.2022.104987.

[32] M. Vafaeefar, K.M. Moerman, T.J. Vaughan, LatticeWorks: An open-source MATLAB toolbox for nonuniform, gradient and multi-morphology lattice generation, and analysis, Mater Des 250 (2025) 113564. https://doi.org/10.1016/J.MATDES.2024.113564.

[33] M. Vafaeefar, H. Katoozian, Finite element-based design, shape and structural rigidity optimization of hip prosthesis, Amirkabir University of Technology, 2018. https://aut.ac.ir/cv/2249/HAMID REZA Katoozian?slc_lang=en/.

[34] B. Mathai, S. Gupta, The influence of loading configurations on numerical evaluation of failure mechanisms in an uncemented femoral prosthesis, Int J Numer Method Biomed Eng 36 (2020) 1–16. https://doi.org/10.1002/cnm.3353.

[35] B. Mathai, S. Dhara, · Sanjay Gupta, Orthotropic bone remodelling around uncemented femoral implant: a comparison with isotropic formulation, Biomech Model Mechanobiol 20 (2021) 1115–1134. https://doi.org/10.1007/s10237-021-01436-6.

[36] H. Si, TetGen, a Delaunay-Based Quality Tetrahedral Mesh Generator, ACM Trans. Math. Softw. 41 (2015). https://doi.org/10.1145/2629697.

[37] H. Si, TetGen - A Quality Tetrahedral Mesh Generator and Three-Dimensional Delaunay Triangulator, Control (2006) 62. http://tetgen.berlios.de/.

[38] C. Quinn, A. Kopp, T.J. Vaughan, A coupled computational framework for bone fracture healing and long-term remodelling: Investigating the role of internal fixation on bone fractures, Int J Numer Method Biomed Eng 38 (2022) e3609. https://doi.org/10.1002/CNM.3609.

[39] R. Huiskes, H. Weinans, H.J. Grootenboer, M. Dalstra, B. Fudala, T.J. Slooff, Adaptive bone-remodeling theory applied to prosthetic-design analysis, J Biomech 20 (1987) 1135–1150. https://doi.org/10.1016/0021-9290(87)90030-3.

[40] R.B. Martin, Porosity and specific surface of bone., Crit Rev Biomed Eng 10 (1984) 179–222.

[41] B.A. Behrens, I. Nolte, P. Wefstaedt, C. Stukenborg-Colsman, A. Bouguecha, Numerical investigations on the strain-adaptive bone remodelling in the periprosthetic femur: Influence of the boundary conditions, Biomed Eng Online 8 (2009) 1–9. https://doi.org/10.1186/1475-925X-8-7.

[42] ABAQUS, Dassault Systèmes, (2019).

[43] J.W. Nicholson, Titanium Alloys for Dental Implants: A Review, Prosthesis 2 (2020) 100–116. https://doi.org/10.3390/prosthesis2020011.

[44] D. Li, W. Liao, N. Dai, Y.M. Xie, Comparison of Mechanical Properties and Energy Absorption of Sheet-Based and Strut-Based Gyroid Cellular Structures with Graded Densities, Materials 12 (2019) 2183. https://doi.org/10.3390/ma12132183.

[45] L.C. Gerhardt, A.R. Boccaccini, Bioactive glass and glass-ceramic scaffolds for bone tissue engineering, Materials 3 (2010) 3867–3910. https://doi.org/10.3390/ma3073867.

[46] E.F. Morgan, G.U. Unnikrisnan, A.I. Hussein, Bone Mechanical Properties in Healthy and Diseased States, Annual Review OfBiomedical Engineering 20 (2018). https://doi.org/10.1146/annurev-bioeng-062117-121139.

[47] M.O. Heller, G. Bergmann, J.P. Kassi, L. Claes, N.P. Haas, G.N. Duda, Determination of muscle loading at the hip joint for use in pre-clinical testing, J Biomech 38 (2005) 1155–1163. https://doi.org/10.1016/J.JBIOMECH.2004.05.022.

[48] M.J. Chashmi, A. Fathi, M. Shirzad, R.A. Jafari-Talookolaei, M. Bodaghi, S.M. Rabiee, Design and analysis of porous functionally graded femoral prostheses with improved stress shielding, Designs (Basel) 4 (2020) 1–15. https://doi.org/10.3390/designs4020012.

[49] J. Nandy, H. Sarangi, S. Sahoo, A Review on Direct Metal Laser Sintering: Process Features and Microstructure Modeling, Lasers in Manufacturing and Materials Processing 6 (2019) 280–316. https://doi.org/10.1007/S40516-019-00094-Y/TABLES/4.

[50] M. Walczak, M. Szala, Effect of shot peening on the surface properties, corrosion and wear performance of 17-4PH steel produced by DMLS additive manufacturing, Archives of Civil and Mechanical Engineering 21 (2021) 1–20. https://doi.org/10.1007/S43452-021-00306-3/FIGURES/13.

[51] G. Motomura, N. Mashima, H. Imai, A. Sudo, M. Hasegawa, H. Yamada, M. Morita, N. Mitsugi, R. Nakanishi, Y. Nakashima, Effects of porous tantalum on periprosthetic bone remodeling around metaphyseal filling femoral stem: a multicenter, prospective, randomized controlled study, Scientific Reports 2022 12:1 12 (2022) 1–9. https://doi.org/10.1038/s41598-022-04936-2.

[52] N. Tan, R.J. van Arkel, Topology Optimisation for Compliant Hip Implant Design and Reduced Strain Shielding, Materials 2021, Vol. 14, Page 7184 14 (2021) 7184. https://doi.org/10.3390/MA14237184.

[53] H. Mehboob, F. Ahmad, F. Tarlochan, A. Mehboob, S.H. Chang, A comprehensive analysis of bio-inspired design of femoral stem on primary and secondary stabilities using mechanoregulatory algorithm, Biomech Model Mechanobiol 19 (2020) 2213–2226. https://doi.org/10.1007/s10237-020-01334-3.

[54] F. Bartolomeu, M.M. Costa, N. Alves, G. Miranda, F.S. Silva, Additive manufacturing of NiTi-Ti6Al4V multi-material cellular structures targeting orthopedic implants, Opt Lasers Eng 134 (2020) 106208. https://doi.org/10.1016/J.OPTLASENG.2020.106208.

[55] Zimmer Biomet, Arcos ® One-piece Femoral Revision System, 2016.

[56] D.F. Amanatullah, H. Siman, G.D. Pallante, D.B. Haber, R.J. Sierra, R.T. Trousdale, Revision total hip arthroplasty after removal of a fractured well-fixed extensively porouscoated femoral component using a trephine, Bone and Joint Journal 97-B (2015) 1192–1196. https://doi.org/10.1302/0301-620X.97B9.35037.

[57] B.A. Masri, P.A. Mitchell, C.P. Duncan, Removal of Solidly Fixed Implants During Revision Hip and Knee Arthroplasty, Journal of the American Academy of Orthopaedic Surgeons 13 (2005) 18–27.

[58] P. Müller, A. Synek, T. Stauß, C. Steinnagel, T. Ehlers, P.C. Gembarski, D. Pahr, R. Lachmayer, Development of a density-based topology optimization of homogenized lattice structures for individualized hip endoprostheses and validation using micro-FE, Scientific Reports 2024 14:1 14 (2024) 1–14. https://doi.org/10.1038/s41598-024-56327-4.

[59] A. Bandyopadhyay, I. Mitra, S.B. Goodman, M. Kumar, S. Bose, Improving biocompatibility for next generation of metallic implants, Prog Mater Sci 133 (2023) 101053. https://doi.org/10.1016/J.PMATSCI.2022.101053.

[60] A.T. Sidambe, J.K. Oh, Biocompatibility of Advanced Manufactured Titanium Implants—A Review, Materials 2014, Vol. 7, Pages 8168-8188 7 (2014) 8168–8188. https://doi.org/10.3390/MA7128168.

[61] N.S. Manam, W.S.W. Harun, D.N.A. Shri, S.A.C. Ghani, T. Kurniawan, M.H. Ismail, M.H.I. Ibrahim, Study of corrosion in biocompatible metals for implants: A review, J Alloys Compd 701 (2017) 698–715. https://doi.org/10.1016/J.JALLCOM.2017.01.196.